\documentclass[12pt]{article}
\usepackage{amsmath, amssymb, bm}
\usepackage{color}
\usepackage{graphicx}
\usepackage{placeins}
\usepackage{jheppub}
\usepackage{hyperref}
\usepackage{cleveref}
\begin{document}

\title{Committing to Bubbles: Finding the Critical Configuration on the Lattice}

\author[a]{Tomasz P. Dutka}

\affiliation[a]{School of Physics, Korea Institute for Advanced Study, Seoul, 02455, Republic of Korea}

\emailAdd{tdutka@kias.re.kr}

\abstract{The nucleation of bubbles in first-order phase transitions is traditionally characterised by the critical bubble: defined as the saddle-point solution of the Euclidean action that separates collapsing from expanding field configurations. While this picture is exact in the noiseless, zero-temperature limit, thermal fluctuations introduces stochasticity which can influence the behaviour of the field configuration. In this work, we develop a purely statistical criterion for identifying the critical bubble by leveraging the concept of the ``committor'' probability: the likelihood that a given local field configuration evolves to the true vacuum before returning to the false vacuum. Using ensembles of lattice simulations with controlled thermal noise, we extract the committor probability during the evolution of a bubble from sub- to super-criticality. We find this approach to be robust, accounts for finite-temperature effects, and allows independent verification of bounce-based predictions. To demonstrate this, we compare the average profile obtained via the committor probability method to standard theory for a given model and find strong agreement, particularly at the core of the bubble. Importantly, we also observe that the behaviour of the committor probability with time is smooth and well defined. This method establishes a robust, simulation-driven framework for studying nucleation dynamics in thermal field theories and may be especially applicable in cases where analytical control might be limited.}
\maketitle

\section{Introduction}

Cosmological first-order phase transitions are hypothetical phenomena that arise naturally in many extensions of the Standard Model (SM). In the early universe, symmetries that are spontaneously broken at low temperatures are typically restored at high temperatures, and as the universe cools during expansion the transition back to the broken phase can proceed discontinuously through bubble nucleation. Although the SM itself does not predict a first-order transition, its open problems motivate new physics that can. Modifications of the Higgs sector or the introduction of new symmetries at higher scales can render the electroweak transition or other transitions in the thermal history first-order. The dynamics of such transitions are not only of intrinsic interest but also have direct implications for unresolved questions in cosmology and particle physics. Bubble expansion can 
generically: (i) produce sufficiently strong stochastic gravitational wave signals\,\cite{Witten:1984rs,Hogan:1986qda,Kosowsky:1991ua,Kosowsky:1992vn,Kamionkowski:1993fg,Espinosa:2010hh,Caprini:2019egz, Schmitz:2020syl,Athron:2023xlk}, (ii) 
affect the baryon asymmetry or the dark-matter abundance~\cite{
Chway:2019kft, Baker:2019ndr, Ahmadvand:2021vxs, Hong:2020est, Azatov:2021ifm, Asadi:2021pwo,
Shaposhnikov:1987tw, Cline:2018fuq, Hall:2019ank, Fujikura:2021abj, Baldes:2021vyz, Azatov:2021irb, Arakawa:2021wgz, Huang:2022vkf, Dasgupta:2022isg, Chun:2023ezg}, (ii) 
can provide mechanisms for primordial black-hole formation distinct from inflationary scenarios~\cite{
Hawking:1982ga, Moss:1994pi, Jung:2021mku,
Sato:1981bf, Maeda:1981gw, Kodama:1982sf, Hall:1989hr, Kusenko:2020pcg,
Khlopov:1999ys, Lewicki:2019gmv,
Gross:2021qgx, Baker:2021nyl, Kawana:2021tde, Baker:2021sno,
Liu:2021svg, Hashino:2021qoq, Huang:2022him, DeLuca:2022bjs, Kawana:2022olo, Lewicki:2023ioy, Gouttenoire:2023naa, Jinno:2023vnr, Ai:2024cka},
and, (iv) if strongly supercooled, can dilute relics that otherwise overclose the Universe~\cite{Lyth:1995hj, Lyth:1995ka, Barreiro:1996dx, Jeong:2004hy, Easther:2008sx}.

The dynamics of first-order phase transitions in quantum field theory are governed by the nucleation of bubbles of the stable phase within a surrounding metastable environment. The central theoretical object in this process is the critical bubble: the field configuration at which the free energy (or Euclidean action) is stationary but not minimal, representing the saddle point that separates shrinking (subcritical) from expanding (supercritical) bubbles. In the presence of fluctuations, thermal or quantum, this configuration marks the point where a fluctuation has equal probability to collapse back into the false vacuum or to expand and convert the surrounding space into the true vacuum. In conventional treatments~\cite{Coleman:1977py,Callan:1977pt,PhysRevLett.46.388,Linde:1981zj,Linde:1980tt,Langer:1967ax, Langer:1969bc, Langer:1974cpa, Kramers:1940zz, LINDE198137, 1983544}, the critical bubble is identified with the so-called bounce solution of the Euclidean field equations: an $\mathcal{O}(D)$-symmetric, static saddle point of the action. This solution serves as the ``separatrix'' between decaying and surviving field trajectories.

This picture, however, rests on the assumption of an idealised deterministic evolution governed solely by the Euclidean action and its equations of motion: a noiseless limit. In realistic finite-temperature settings, thermal fluctuations introduce stochastic noise on top of the deterministic roll of the field. If the phase-transitioning field couples strongly to the thermal plasma, it is not unreasonable that configurations that appear to lie beyond the separatrix may still be driven back toward the false vacuum by a sequence of thermal kicks, while profiles that would na\"ively collapse may be rescued by noise and pushed over the barrier into the true vacuum. Such effects may complicate the definition of the critical bubble.

This motivates a purely statistical definition of the critical bubble. Rather than relying on a single analytic saddle point, one can define the critical bubble in terms of probabilistic criteria applied to ensembles of fluctuating configurations. Concretely, given a field profile evolved under Langevin dynamics with thermal noise, one measures the probability of expansion versus collapse across many stochastic realisations for a given field profile as an initial condition: the committor probability. The critical bubble is then defined statistically as the profile for which these two outcomes occur with equal frequency. 

Such a definition of criticality has several advantages. First, it provides an independent verification of the standard bounce picture, allowing a direct comparison between the statistical criterion and the conventional saddle-point solution using lattice simulations. Second, it reveals how thermal fluctuations may modify the na\"ive separatrix behaviour. It allows for both the field, $\phi$, and its momentum profile, $\pi$, to be analysed at the point of criticality. Finally, it yields a robust, simulation-based framework that does not presuppose a perfectly symmetric solution, and can thus accommodate more complicated situations where inhomogeneities, noise, or model-specific dynamics obstruct analytic control.

Related ideas have also been explored from a complementary viewpoint in the \emph{overdamped limit} of Langevin dynamics, where the critical configuration is defined independently of the explicit stochastic noise.
In that approach, the transition surface coincides with the static separatrix of the overdamped system, providing a clean deterministic definition of criticality \cite{Hirvonen:2024rfg,Hirvonen:2025hqn}.
Our statistical criterion should be fully consistent with this construction in the limit $\eta \rightarrow \infty$, where inertia and noise become negligible.
At finite~$\eta$, however, the two definitions differ slightly: using the physical damping in the field equation allows us to study how stochastic kicks and finite-time dynamics deform the idealised separatrix.
In this sense, the committor–probability framework can be viewed as a natural extension of the overdamped definition to realistic thermal settings.

In this work we seek to explore how robustly the committor probability can be applied to lattice studies of first-order phase transitions. We develop and implement such a statistical criterion for the critical bubble and evaluate how robust such a treatment can be when applied to lattice simulations of first order phse transitions. Using ensembles of lattice simulations (see e.g.~\cite{Rummukainen:1998as,Moore:2001vf,Hiramatsu:2014uta,Gould:2024chm,Hirvonen:2025hqn}) with controlled thermal noise, we establish quantitative measures for identifying the critical profile. Our results show that the statistical definition is well defined (and evolves in a smooth way in time), provides a natural generalisation of the separatrix concept and, offers a practical tool for testing and refining theoretical predictions of nucleation rates.

This paper is organised as follows. \Cref{sec:2} briefly discusses the bounce solution in phase transitions as well as introduces the idea of the committor probability.
\Cref{sec:3} describes simulation methodology and our method of extracting the committor probability, and from this the critical bubble profile, from lattice simulations. 
\Cref{sec:4} demonstrates the extracted committor probability and critical bubble profiles in the case of a specific thermal potential.
Finally, in~\Cref{sec:5} we conclude.

\section{Theoretical background}
\label{sec:2}

A useful analogy in this work comes from statistical physics and chemical reaction theory, where the concept of a committor probability~\cite{annurev:/content/journals/10.1146/annurev.physchem.53.082301.113146} is central. For a given microscopic configuration, the committor probability is defined as the likelihood that the system, when evolved with thermal noise, will first reach the “product” state rather than returning to the “reactant” state. The dividing surface between the two basins is defined as the set of configurations with committor probability equal to one half. 
In our context, the false vacuum plays the role of the reactant, the true vacuum the product, and the critical bubble is precisely the configuration with committor probability $p_B = 1/2$.

In the context of first–order phase transitions, the \emph{separatrix} is the codimension-1 hypersurface in configuration space that divides configurations which relax back to the metastable (false vacuum) from those which evolve to the stable (true vacuum). The \emph{critical bubble} lies precisely on this separatrix. It is a saddle–point solution of the equations of motion or, equivalently, of the free–energy functional. In the noiseless limit, configurations with sufficiently small amplitude or spatial extent tend to collapse, while sufficiently large ones tend to grow.

Consider a scalar field in $d+1$ dimensions with Euclidean action
\begin{equation}
S_E[\phi] = \int d\tau\, d^d x \left[ \tfrac{1}{2}(\partial_\tau \phi)^2 + \mathcal{S}_d \right],
\end{equation}
where the spatial action density is
\begin{equation}
\mathcal{S}_d \equiv \tfrac{1}{2} (\nabla \phi)^2 + \Delta V_{\rm eff}(\phi;T).
\end{equation}
The effective potential $V_{\rm eff}(\phi;T)$ has a metastable minimum $\phi_{\rm FV}$ (false vacuum) and a stable minimum $\phi_{\rm TV}$ (true vacuum). At zero temperature, $V_{\rm eff}(\phi;0)\equiv V(\phi)$.

At finite temperature, Euclidean time is compact, $\tau \sim \tau+\beta$. In the high-$T$ limit, bounce solutions are driven by thermal escape, static in $\tau$, and the action factorises as
\begin{equation}
S_E[\phi] = \beta \int d^d x\, \mathcal{S}_d.
\end{equation}

The \textit{stationary} bounce solution satisfies
\begin{equation}
\phi''(\rho) + \frac{D-1}{\rho}\,\phi'(\rho) = V_{\rm eff}'(\phi;T),
\qquad 
\phi'(0)=0, \quad \phi(\rho \to \infty)=\phi_{\rm FV},
\end{equation}
with $\mathcal{O}(D)$ symmetry. Here
\[
\rho \equiv 
\begin{cases}
\sqrt{\tau^2 + |\mathbf{x}|^2}, & \text{quantum tunnelling ($T=0$, $D=d+1$)}, \\[6pt]
|\mathbf{x}|, & \text{thermal tunnelling ($T>0$, $D=d$)}.
\end{cases}
\]

\subsection{Deterministic vs.~thermal separatrix}

At zero temperature, field configurations evolve deterministically via
\begin{equation}
\ddot{\phi}(\mathbf{x},t) - \nabla^2 \phi(\mathbf{x},t) + V'(\phi) = 0.
\end{equation}
In this limit, the separatrix is sharp: it coincides with the unstable critical bubble profile $\phi_{\rm crit}(r)$. Perturbations below this profile, $\phi_{\rm crit}(r) - \delta\phi$, collapse to the false vacuum, while perturbations above it,$\phi_{\rm crit}(r)+\delta\phi$, grow toward the true vacuum. The outcome is fully determined by the initial condition.

In the presence of a thermal bath, however, the field instead evolves stochastically, e.g. under Langevin dynamics:
\begin{equation}
\ddot{\phi} - \nabla^2 \phi + \eta \dot{\phi} + V'(\phi) = \xi(\mathbf{x},t),
\end{equation}
which has been used frequently in the past to model thermal phase transitions~\cite{Farakos:1994xh, Borrill:1994nk, Yamaguchi:1996dp, Borrill:1996uq, Cassol-Seewald:2007oak, Hiramatsu:2014uta, Gould:2024chm, Pirvu:2024ova, Pirvu:2024nbe}. Here $\eta$ is a damping coefficient and $\xi$ is white Gaussian noise\footnote{The escape rate per unit volume can be obtained from the Fokker–Planck equation, which is equivalent to the Langevin formulation, as demonstrated in~\cite{Berera:2019uyp, Gould:2021ccf}.}. In the overdamped limit this reduces to
\begin{equation}
\eta \dot{\phi} = -\frac{\delta S_E}{\delta \phi} + \xi.
\end{equation}
Thermal fluctuations can therefore alter the deterministic picture: a configuration just below the critical bubble may be kicked over the barrier and expand, while a configuration just above it may be pushed back and collapse.

For such stochastic systems, we will define the separatrix statistically: for a given initial configuration $\phi_0$, the probability that the system reaches the true vacuum before the false vacuum is
\begin{equation}
p_{\rm B}[\phi_0] \;\equiv\; 
\mathbb{P}\!\left( \phi(t) \rightarrow \phi_{\rm TV}\;\;\text{before}\;\; \phi(t)\rightarrow \phi_{\rm FV} \,\big|\, \phi(0)=\phi_0 \right).
\end{equation}
The \emph{statistical critical bubble} is then defined as
\begin{equation}
\phi_{\rm crit}: \quad p_{\rm B}[\phi_{\rm crit}] = \tfrac{1}{2}.
\end{equation}
In the noiseless limit, $T \to 0$, the committor probability $p_{\rm B}$ takes only the values $0$ or $1$, with a sharp discontinuity at the critical bubble. In this sense the statistical criterion reduces to the deterministic bounce solution. In a noisy environment however, one expects that $p_{\rm B}$ will evolve in time, smoothly interpolating between $0$ and $1$ as the fluctuation grows to macroscopic size.

\section{Simulation Methodology}
\label{sec:3}

For our simulations, we consider a $\mathbb{Z}_2$-symmetric scalar potential, where the zero-temperature vacuum expectation value 
of the scalar field $\phi$
spontaneously breaks the $\mathbb{Z}_2$ symmetry where $\phi \to -\phi$. 
Then as the temperature decreases, the finite-temperature effective potential $V_T(\phi)$ induces a phase transition from $\phi_{\rm FV} = 0$ to $|\phi|= \phi_{\rm TV} \equiv \phi_{\rm vev} >0$. 
Building on past lattice simulations~\cite{Dutka:2025oqt}\footnote{See~\cite{Batini:2023zpi} as an alternative example of employing real-time lattice dynamics of bubble nucleation in $3+1$ dimensions.}, and due to past familiarity, we choose to simulate the following effective potential on the lattice
\begin{equation}
\label{eq:4d-pot}
    V_{T}(\phi) = \frac{1}{2}m_T^2 \phi^2 - \frac{\lambda_T}{4!} \phi^4 + \frac{\epsilon}{6!} \phi^6.
\end{equation}
This type of potential can naturally arise in supercooled phase transition theories, such as some SUSY theories with a tachyonic instability stabilised by higher-order terms.
The parameters in~\cref{eq:4d-pot}
should be understood as simulation parameters with thermal effects included, as the zero-temperature theory will have negative curvature at the origin.

The renormalisable parameters, $m_T$ and $\lambda$, effectively vary the bounce action, $S_3/T$, while $\epsilon$ adjusts the potential energy difference $\Delta V_T$ between the true and false vacua.
We fix the parameters to similar values studied in the past~\cite{Dutka:2025oqt} and shown in~\cref{tab:c_H}, simply for convenience as it is already known that the nucleation rate is sufficiently large such that bubbles will spontaneously nucleate in reasonable simulation times. 
We leave the possibility of comparing the extracted critical bubble profiles for different parameter choices, or even different potentials: e.g. $\kappa_2 \phi^2 - \kappa_3 \phi^3 + \kappa_4 \phi^4$, to the future.

The extrema of~\cref{eq:4d-pot} are given by
\begin{equation}
    \frac{\partial V_T}{\partial \phi} = 0 \implies \phi \in \{0,\,\phi_{\mathrm{top}},\,\phi_{\mathrm{vev}}\}
\end{equation}
where $\phi_{\mathrm{top}} = \phi_b/\sqrt{2}$, $\phi_b^2 \simeq (12/\lambda_T) m_T^2$ and $\phi_{\mathrm{vev}} \simeq 2\sqrt{5}\sqrt{\lambda_T/\epsilon}$, provided that $\epsilon$ is small: $\epsilon \phi_b^2 \ll \lambda_T$.

Following the literature~\cite{Farakos:1994xh, Borrill:1994nk, Yamaguchi:1996dp, Borrill:1996uq, Cassol-Seewald:2007oak, Hiramatsu:2014uta, Gould:2024chm, Pirvu:2024ova, Pirvu:2024nbe}, the dynamics of the scalar field $\phi$ 
in the presence of thermal fluctuations
can be modeled through the 
the Langevin equation\footnote{In principle a multiplicative noise term $\xi_m(\mathbf{x}) \phi(\mathbf{x})$ should also be included but appears only to affect the time scale for the system to equilibrate. This, however, will complicate the implementation of the simulation~\cite{Cassol-Seewald:2007oak} so we neglect this contribution for simplicity.}:
\begin{equation}
\partial_t^2 \phi({\bf x}, t) + \eta \partial_t \phi({\bf x}, t) - \nabla^2 \phi({\bf x}, t) + \partial_\phi V_T(\phi) = \xi({\bf x}, t),
\label{eq:sec2langevin}
\end{equation}
where \( \xi({\bf x}, t) \) represents a stochastic noise term satisfying the following statistical properties:
\begin{equation}
\langle \xi({\bf x}, t) \rangle = 0, \quad \langle \xi({\bf x}, t) \xi({\bf x}', t') \rangle = D \delta^{(3)}({\bf x} - {\bf x}') \delta(t - t'), 
\quad D = 2T\eta. 
\label{eq:sec2fluc-diss}
\end{equation}
$\eta$ is understood as a damping coefficient and $\xi$ as uncorrelated, white noise.
Such an assumption will remain valid as long as 
the lattice spacing of the simulation
is larger than the spatial and temporal correlation lengths of the noise, generated by the fermions and bosons within the thermal bath. 
This is  typically $\ell \sim (\pi T)^{-1}$ for fermions and exponentially damped for bosons~\cite{Yamaguchi:1996dp}.
The fluctuation-dissipation relation, $D=2T\eta$, 
ensures that the equilibrium values for $\phi$ do not depend on the damping parameter $\eta$, which only serves to control the time scale to equilibration. 

We simulate\footnote{Code from which this analysis was based is available at \href{https://github.com/tpdutka/latbubble}{\url{https://github.com/tpdutka/latbubble}}} a 3-dimensional effective thermal field theory containing the light bosonic field, the zero Matsubara mode, which we write as $\phi_3$, with the remaining heavy fields integrated out~\cite{Farakos:1994xh}. As the non-zero modes scale as $\pi n T$, this requires sufficiently high temperatures. 
The parameters within the dimensionally reduced theory are related to the $3 + 1$ dimensional thermal theory through powers of temperature. In our case,
\begin{align}
    m_3^2 = m_T^2,\quad \lambda_3 = \lambda_T T \quad   \text{and} \quad  \epsilon_3 = \epsilon T^2
\end{align}
which carry appropriate mass dimensions. 
In our numerical simulations, we express all dimensionsful parameters relative to temperature, i.e. setting $T=1$. Importantly for the results shown later, lattice renormalisation counterterms are required in order to match lattice theories to continuum ones. Most importantly in this case is the mass renormalsation counter term, $m_{\rm lat}^2 = m_3^2 + \delta m^2$, with $\delta m^2$ scaling inversely with lattice spacing and can dominate the value of $m_{\rm lat}^2$. These counterterms are described in~\cref{app:A}.

Following the standard high-temperature EFT picture, we assume that short-wavelength modes of the thermal bath are in equilibrium and can be integrated out, leaving the Matsubara zero mode of $\phi$ evolved via Langevin dynamics with thermal effective potential $V_{\rm eff}(\phi;T)$. 
A necessary condition is that nonzero Matsubara modes remain heavy relative to the local curvature,
\begin{equation}
    \sqrt{|V''_{\rm eff}(\phi;T)|} \;\ll\; 2\pi T \quad \text{in the field domain relevant for nucleation},
\end{equation}
so that time-dependent Euclidean modes are suppressed and the bounce is static in $\tau$.

For the toy potential of~\cref{eq:4d-pot}, $V(\phi)\sim -\lambda_T \phi^4$ for sufficiently large $|\phi|$, implying that the curvature $\sqrt{|V''_{\rm eff}|}$ will eventually exceed $2\pi T$ beyond some cutoff field value $\phi_{\rm cutoff}$ defined by
\begin{equation}
    \sqrt{|V''_{\rm eff}(\phi_{\rm cutoff};T)|}=2\pi T.
\end{equation}
Since nucleation physics is controlled by the barrier region, the static description is reliable provided the critical profile satisfies $\max_{r \in \mathcal{D}} |\phi(r)| \ll \phi_{\rm cutoff}$. 
In the simulations discussed below, this condition holds; see~\cref{tab:c_H}. 
Numerically, we have also verify that the lattice spectra are dominated by long-wavelength modes during nucleation.

At very large $|\phi|>\phi_{\rm cutoff}$ values, well past critical bubble formation, additional dynamics (e.g.\ nonzero Matsubara modes, $\phi$-dependent damping, and hydrodynamic backreaction) should be considered and would require an extended treatment.

\subsection{Benchmark and nucleation estimates}
\label{sec:Benchmarks}

In the tachyonic limit $\epsilon\to 0$ of~\cref{eq:4d-pot}, the thermally induced action admits the semi-analytic estimate~\cite{Linde:1981zj}
\begin{equation}
\label{Eq:S3overT_simulation}
    \frac{S_3}{T}\;\simeq\; \frac{114\,\sqrt{m^2}}{\lambda\,T}\,,
\end{equation}
which remains valid when the true minimum lies well beyond the barrier. 
For small barriers, which happens to be the case for the parameters in~\cref{tab:c_H}, the prefactor in $\Gamma_n \simeq A\, e^{-S_3/T}$ can contribute appreciably; to get an idea of its minimum size, we evaluate it numerically with \texttt{BubbleDet}~\cite{Ekstedt:2023sqc}.

\begin{table}[t]
    \centering
    \renewcommand{\arraystretch}{1.15}
\begin{tabular}{|c|c||c|c||c|c||c |c |}
    \hline
    & \{$\lambda$, $m_T$, $\epsilon$\} & $\phi_{\rm vev}/T$ & $\left|\Delta V_T/T^4\right|$ & $S_3/T$ & $\log \Gamma_n/T^4$ & $\frac{\sqrt{|V''(\phi_c)|}}{(2\pi T)}$ & $\frac{\phi_{\rm cutoff}}{\phi_{\rm c}}$\\
    \hline
    A & \{$2$, $0.122\, T$, $T^2/100$\} & $63.26$ & $4.44 \times 10^5$ & $6.98$ & $-15.6$ & $0.145$ & $6.889$\\
    \hline 
\end{tabular}
    \caption{Benchmark parameters and nucleation estimates. 
    The last two columns quantify the validity of the dimensionally reduced description at the center of the critical bubble ($\phi_c$) and the distance to the cutoff $\phi_{\rm cutoff}$ defined by $\sqrt{|V''(\phi_{\rm cutoff})|}=2\pi T$. The estimates for $S_3$ and $\Gamma$ are obtained using \texttt{CosmoTransitions}~\cite{Wainwright:2011kj} and \texttt{BubbleDet}~\cite{Ekstedt:2023sqc}.}
\label{tab:c_H}
\end{table}

For the exploratory study here we choose $\lambda=2$ so that the expected time to first nucleation,
\begin{equation}
    t_{\rm first\text{-}bubble}\;\sim\;\frac{1}{\Gamma_n\,L^3}\,,
\end{equation}
is short enough to allow many independent realisations. 
For smaller $\lambda$, the scaling $S_3/T \propto \lambda^{-1}$ increases $t_{\rm first\text{-}bubble}$ considerably; conversely, realistic models with $S_3/T=\mathcal{O}(100)$ would require prohibitively long runs unless one seeds bubbles, see e.g.~\cite{Berg:1991cf,Berg:1992qua,Gould:2024chm} or~\cite{Bai:2024pii} for a machine learning approach.

We note that the benchmark in table~\ref{tab:c_H} does not lie in a parametrically weak-coupling regime. In particular, the dimensionless expansion parameter $\lambda_3/(4\pi m_3)\simeq 1.3$ is of order unity. This semi-strong coupling reflects a deliberate compromise: increasing $\lambda$ reduces the nucleation barrier and allows spontaneous bubble formation to be observed on feasible simulation timescales, whereas parametrically smaller couplings would lead to exponentially suppressed nucleation and prohibitively long runs for this initial exploratory study of the separatrix.

In this regime, semiclassical bounce-based predictions should be interpreted primarily as qualitative guidance rather than precision benchmarks. The critical bubble obtained from the semiclassical instanton calculation corresponds to a saddle point of the renormalised effective action within the dimensionally reduced continuum description. By contrast, the separatrix configurations identified in the lattice simulations arise from the full stochastic dynamics and therefore incorporate fluctuation effects nonperturbatively. As a result, deviations between the semiclassical saddle-point profile and the committor-defined critical configurations are not unreasonable at this coupling strength.

It would be valuable in future work to repeat this analysis in more weakly coupled regimes, where the saddle-point approximation becomes parametrically controlled, in order to study the approach to the semiclassical limit.

The simulation time step is chosen smaller than the characteristic oscillation time near the global minimum,
\begin{equation}
    \tau \;\sim\; \frac{1}{\sqrt{V''_{\rm eff}(\phi_{\rm vev};T)}} 
    \;\sim\; \frac{\sqrt{\epsilon}}{\lambda}\,,
\end{equation}
which would ensure stability during post-nucleation roll-down and subsequent oscillations even though the simulations are halted before reaching this stage in this specific case. 

\subsection{Primary simulations and subgrid capture}
\label{sec:PrimarySubgrid}

\begin{figure}[t]
    \centering
    \begin{minipage}{0.45\textwidth}
        \includegraphics[width=\textwidth]{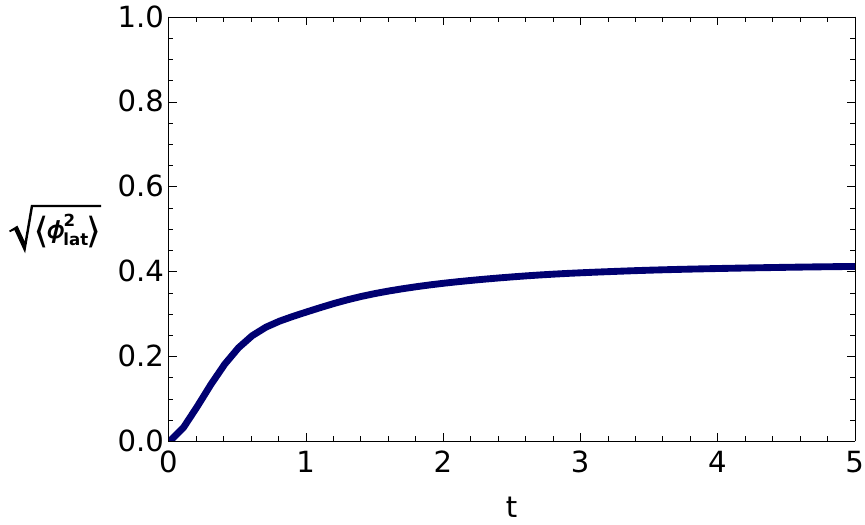}
    \end{minipage}\hfill
    \begin{minipage}{0.45\textwidth}
        \includegraphics[width=\textwidth]{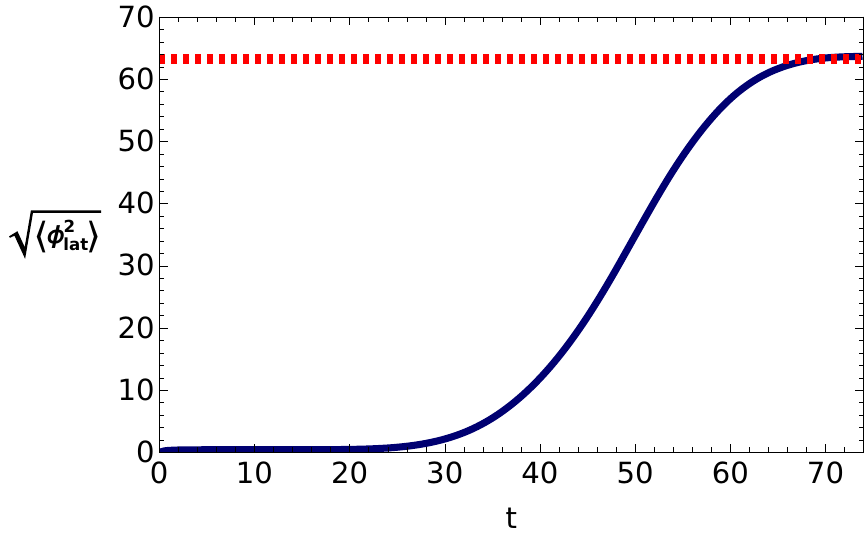}
    \end{minipage}
    \caption{Example evolution for scalar potential considered in this paper when coupled to a thermal bath, shown in~\cref{tab:c_H}. 
    \textbf{(left)} rapid thermalisation around the metastable origin starting from 
    $\phi(\mathbf{x},0)=0=\dot\phi(\mathbf{x},0)$. 
    \textbf{(right)} on a much longer timescale the system transitions to the true vacuum. 
    The red dashed line marks $\phi_{\rm vev}$ of the continuum effective theory which agrees with the lattice simulation due to the inclusion of lattice renormalisation counterterms, such as $Z_\phi$, $\delta \lambda$ and most importantly $\delta m^2$.}
    \label{fig:therm_time}
\end{figure}

We initialise $N_{\rm primary} \sim \mathcal{O}(100)$ primary, large lattice simulations with
\begin{equation}
    \phi(\mathbf{x},0)=0, \qquad \pi(\mathbf{x},0) \equiv\dot\phi(\mathbf{x},0)=0,
\end{equation}
which, though unphysical, has no measurable impact on late-time observables because the system rapidly reaches thermal equilibrium around the metastable minimum. 
The thermalisation timescale is hierarchically shorter than the escape time from the false vacuum, as seen in~\cref{fig:therm_time}: the left panel shows the $O(T)$ rise of $\sqrt{\langle\phi^2\rangle}$ during thermalisation around the origin, whereas the right panel displays the much longer timescale for the full transition.
Time evolution uses the fourth-order symplectic Forest–Ruth integrator~\cite{FOREST1990105} but has been crosschecked with other methods in the past with no significant differences. Details related to the specifics of the lattice simulation can be found in~\cref{app:A}.

\begin{figure}[!t]
    \centering
    \includegraphics[width=\linewidth]{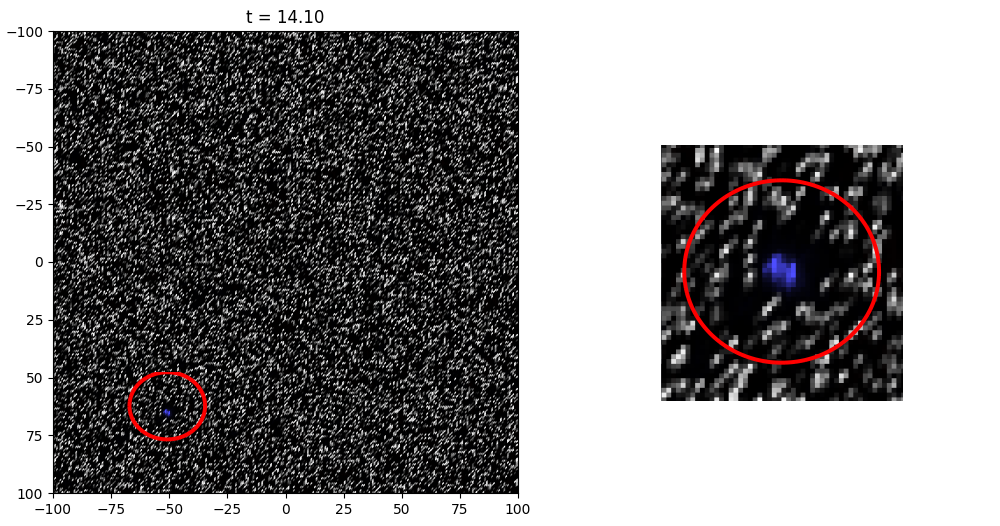}
    \includegraphics[width=0.45\linewidth]{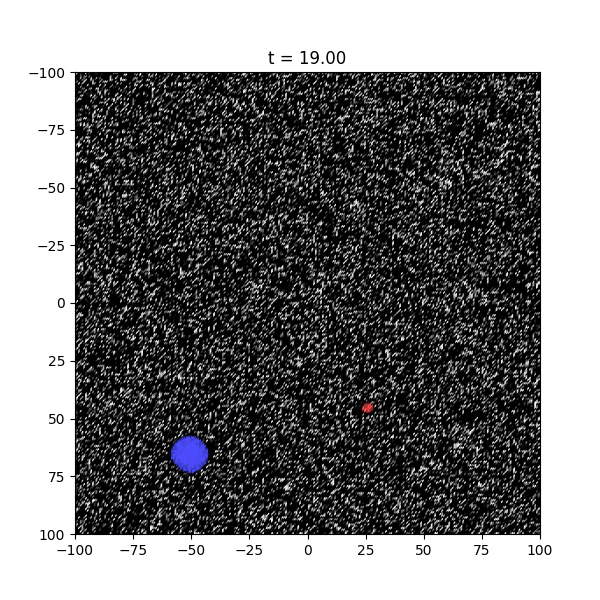}
    \caption{\textbf{(top)} An illustration of the initial simulation procedure. A large lattice volume is simulated until a region of the lattice reaches the true-vacuum position, in this case indicated by blue. Once a small region has formed, this corresponds to a phase transition where this physical region developed a field profile which moved from sub- to super-criticality in time. This data is stored and then used as initial conditions on future, much smaller simulations, to determine at which exact time slice the criticality occurred. \textbf{(bottom)} The bottom plot shows that if the simulation was allowed to continue this region would develop into an expanding bubble including the nucleation of more bubbles at other lattice sites (here shown in red). The simulation procedure closesly follows what was preformed in~\cite{Dutka:2025oqt}.}
    \label{fig:Primary_sim_illustration}
\end{figure}

These primary simulations are run on a volume large compared to the critical radius, $L\gg R_c$, until a localised region transitions, corresponding to bubble formation) 
At first detection of a macroscopic bubble, defined as the core being close to $\phi_{\rm TV}$, the run is halted and all lattice data on a centred subgrid of this core, of size $L_{\rm sub}^3$, are stored for all past time snapshots.
The top part of~\cref{fig:Primary_sim_illustration} shows the primary lattice at the time a large (supercritical) bubble is present, i.e.~well after the critical point separating $\phi_{\rm FV}$ and $\phi_{\rm TV}$, which we use as the end point of the primary simulations. The bottom part of the figure instead corresponds to what would happen if the simulations continued, this bubble begins to expand (with a core at $\phi_{\rm TV}$) and other locations in the large lattice also beginning to nucleate.

Here we choose 
\begin{equation}
    L_{\rm sub} \sim \mathcal{O}(5\text{--}10)\,R_c
\end{equation}
for the size of the subgrid of the data which we save.
This $L_{\rm sub}^3$ sized subgrid data per time step will serve as the initial conditions, $\phi_0(t)$, for subsequent smaller simulations when determining $p_{\rm B}(\phi_0(t))$.

Here, we use results from \texttt{CosmoTransitions}~\cite{Wainwright:2011kj} to get some idea about the size of $R_c$ in order to choose both $L$ and $L_{\rm sub}$ but in principle this could also be determined purely from simulations results for a given potential.
Due to storage limits the full-lattice time history is not retained; only the subgrid data are kept, which suffices because regions far from the bubble remain in a thermal ensemble around the meta-stable phase. 
As the simulation is halted after the first, near-macroscopic bubble, is detected only one sub- to super-critical bubble profile transition is obtained per run.

\subsection{Secondary simulations}

Using the definition of the transition surface and committor probability discussed in the previous section, 
we perform $N_{\rm sub}$ secondary simulations \textit{per primary simulation} for a number of different time steps, $t_i$, with initial conditions extracted from the $L_{\rm sub}^3$ subgrid centred on the bubble found in each primary simulation. Explicitly,
\begin{equation}
    \phi_{\rm sub}(\mathbf{x},0) \equiv \phi_0(t_i)
\end{equation}
for a range of $t_i$ values, and similarly for $\dot{\phi} \equiv \pi$. 

Each secondary simulation is evolved for a fixed duration $t_f$, chosen to be long enough 
to determine whether the configuration grows or collapses, but short enough to avoid 
spontaneous bubble formation unrelated to the seeded profile. In practice we find 
$t_f \sim \mathcal{O}(3$–$5)T$ to be suitable for the specific potential parameters in~\cref{tab:c_H}. The discretisation parameters 
($\Delta t$, $\Delta x$, etc.) are kept identical to those used in the primary simulations.

For each set of initial conditions we perform two sets of $N_{\rm sub}$ simulations:
\begin{enumerate}
    \item A: simulations seeded with the subgrid data from the primary run, to be used to determine the committor probability.
    \item B: \emph{control} simulations in which no data from the primary simulation is inserted and the system is 
    evolved forward with only thermal noise. A short ``prethermalsation'' period is included 
    to ensure comparable statistical properties (e.g.~$\langle \phi^2\rangle$) between the two.
\end{enumerate}
Quite crucially, both A and B runs are evolved with \emph{common random numbers} (CRN) by fixing the random number 
generator seed; the two simulations share identical thermal noise histories at every lattice 
site and time step. The only difference is the presence or absence of the inserted field profile $\phi_0(t)$ initially. 
We therefore interpret any transition to the true vacuum occurring in simulation A to be directly attributable to the initial inserted field configurations if no corresponding transition occurs in simulation B. 
In practice we observe that 
no B simulations nucleate or begin transitioning to $\phi_{\rm TV}$ within $t_f$, confirming that the observed transitions 
in the seeded runs are truly due to the inserted profiles obtained from the primary simulations.

Repeating this procedure $N_{\rm sub}$ times with different random seeds allows us to measure 
the fraction of trajectories that expand to the true vacuum versus those that collapse back 
to the false vacuum. In this work we fix $N_{\rm sub} = 100$ per time step evolved in order to be able to obtain results in a reasonable time frame on a standard desktop computer. This construction is repeated for different primary–simulation snapshots, 
providing a family of initial profiles. The critical bubble is then defined statistically as 
the profile for which the seeded ensemble has equal probability to grow or collapse:
\begin{equation}
    P_{\rm grow} = P_{\rm collapse} = \tfrac{1}{2}.
\end{equation}
Profiles with $P_{\rm grow}<0.5$ are classified as subcritical, while those with 
$P_{\rm grow}>0.5$ are supercritical. As it is not clear which time step, $t_i$, will give a committor probability of one half, extrapolation of $p_{\rm B}(\phi_0(t))$ between different time steps will be used between regions below and above the committor value.

\begin{figure}[t]
    \centering
    \begin{minipage}{0.49\textwidth}
        \includegraphics[width=\textwidth]{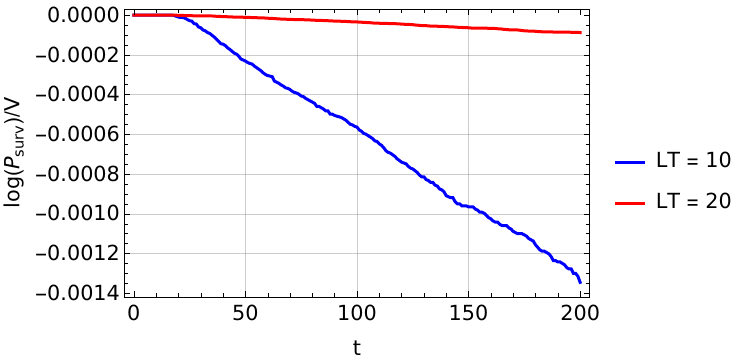}
    \end{minipage}
    \hfill
    \begin{minipage}{0.49\textwidth}
        \includegraphics[width=\textwidth]{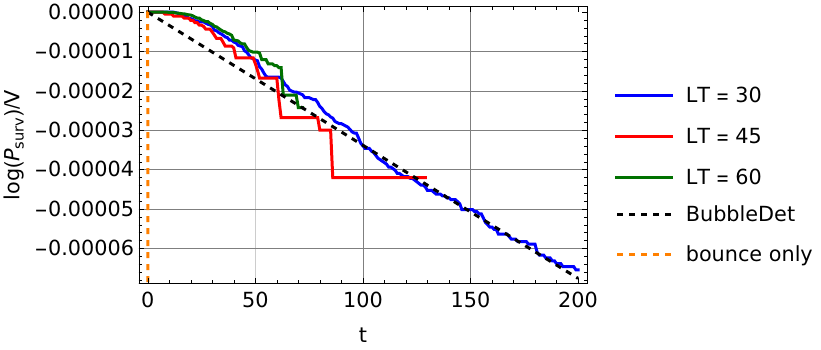}
    \end{minipage}
    \caption{Dependence of the survival probability: the probability of an ensemble of simulations to have nucleated a bubble by time $t$ with different lattice volumes, in units of $T$. 
    The theoretical expectation is $\ln P_{\rm surv}(t)/V = \mathrm{const} - \Gamma t$.
    For small lattice volumes \textbf{(left)} the expected behaviour is not recovered whereas for sufficiently large lattice volumes \textbf{(right)}, the survival probabilities line up. For comparison, semiclassical predictions based on the bounce action are also shown. The curve labeled “BubbleDet” corresponds to the full nucleation rate $\Gamma = A_{\rm BD} \exp(-S_3/T)$, where the prefactor $A_{\rm BD}$ is computed numerically using \texttt{BubbleDet}~\cite{Ekstedt:2023sqc}. The curve labeled “bounce only” shows the exponential suppression factor $\exp(-S_3/T)$ only, which is obtained from \texttt{CosmoTransitions}~\cite{Wainwright:2011kj}. This comparison illustrates the importance of the prefactor contribution to the nucleation rate in this benchmark
    This suggests a minimum lattice volume for the potential in~\cref{tab:c_H} when performing the secondary simulations to determine $p_{\rm B}(\phi_0(t))$ in order to have consistent behaviour.}
    \label{fig:lattice_dimension_effects}
\end{figure}

The physical size $L_{\rm sub}$ must be large enough to avoid finite–volume artefacts but 
small enough for a fast simulation turnaround. As illustrated in~\cref{fig:lattice_dimension_effects}, 
for $L \sim \mathcal{O}(2$–$4)R_c$ the survival probability,  the probability of an ensemble of simulations to have nucleated a bubble by time $t$, deviates from the expected 
$\log(P_{\rm surv}) =\mathrm{const} - \Gamma V t$ behaviour. For $L_{\rm sub}\gtrsim \mathcal{O}(5$–$10)R_c$ 
the correct scaling is observed. The right panel also shows the comparison to semiclassical instanton predictions. Including the numerically computed prefactor from \texttt{BubbleDet}, we obtain $-\log \Gamma \simeq 15.5$, which agrees well with the observed nucleation rate. By contrast, the bounce action alone gives $S_3/T \simeq 7$, highlighting the important role of the prefactor in determining the overall rate in this benchmark. In what follows we adopt $L_{\rm sub}=30T$ as a compromise between 
computational efficiency and physical reliability.

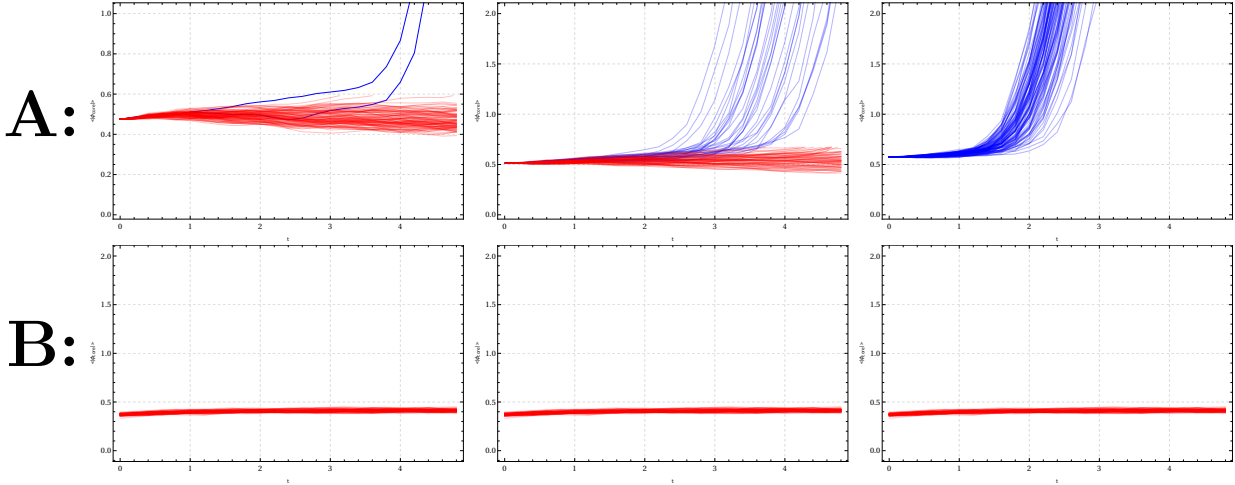
\begin{figure}[t]
    \centering
    \begin{picture}(0,0)
        \put(-215,-43){\makebox(0,0)[r]{\Huge \textbf{A:}}} 
        \put(-215,-130){\makebox(0,0)[r]{\Huge \textbf{B:}}}
    \end{picture}
    
    \hfill
    \begin{minipage}{0.32\textwidth}
        \includegraphics[width=\textwidth]{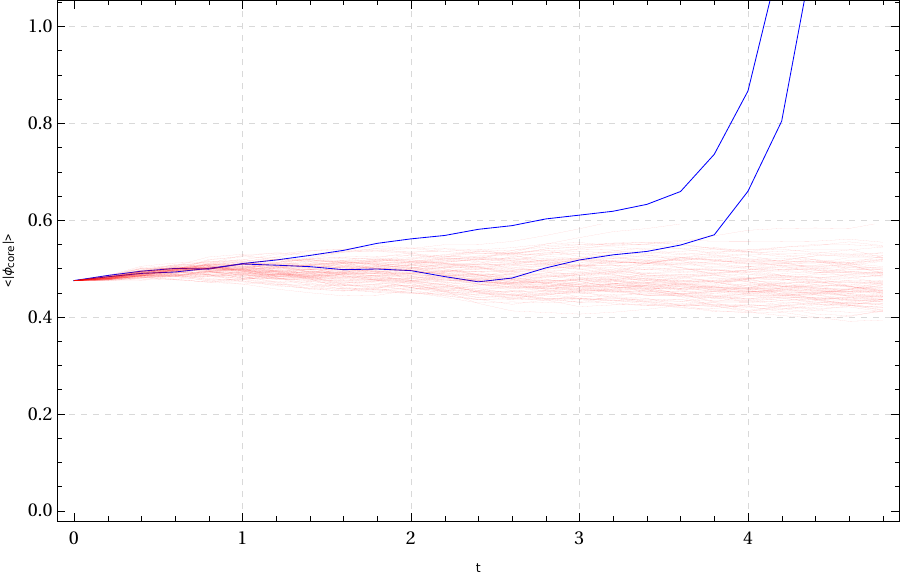}
    \end{minipage}
    \begin{minipage}{0.32\textwidth}
        \includegraphics[width=\textwidth]{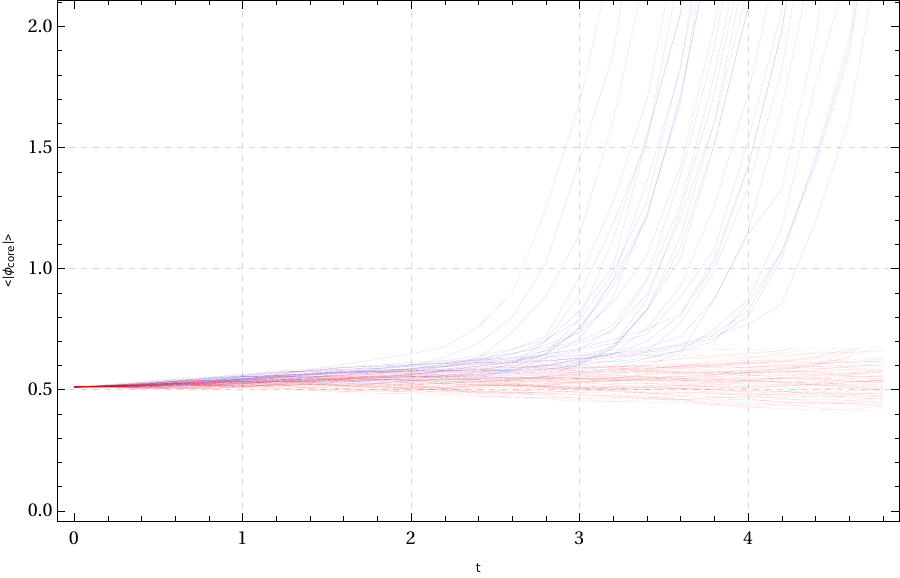}
    \end{minipage}
    \begin{minipage}{0.32\textwidth}
        \includegraphics[width=\textwidth]{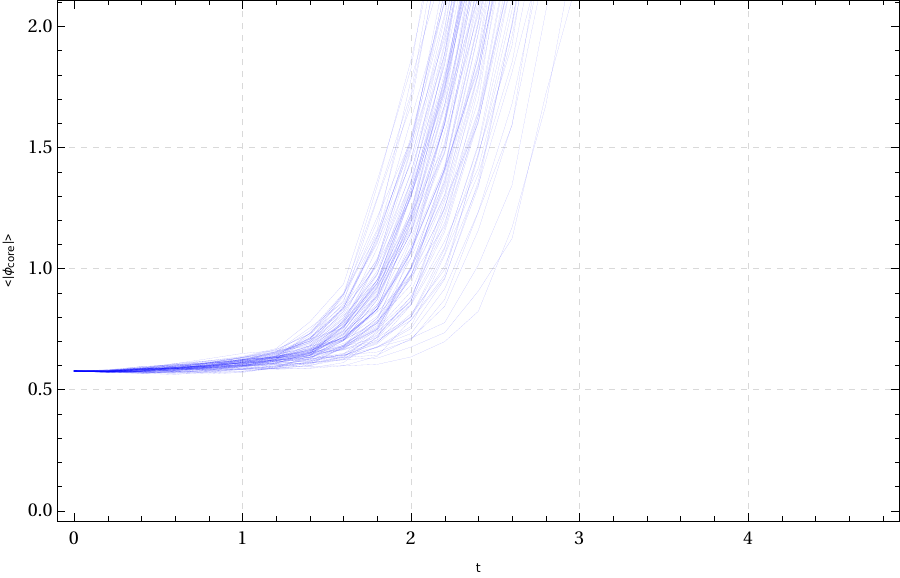}
    \end{minipage}

    \hfill
    \begin{minipage}{0.32\textwidth}
        \includegraphics[width=\textwidth]{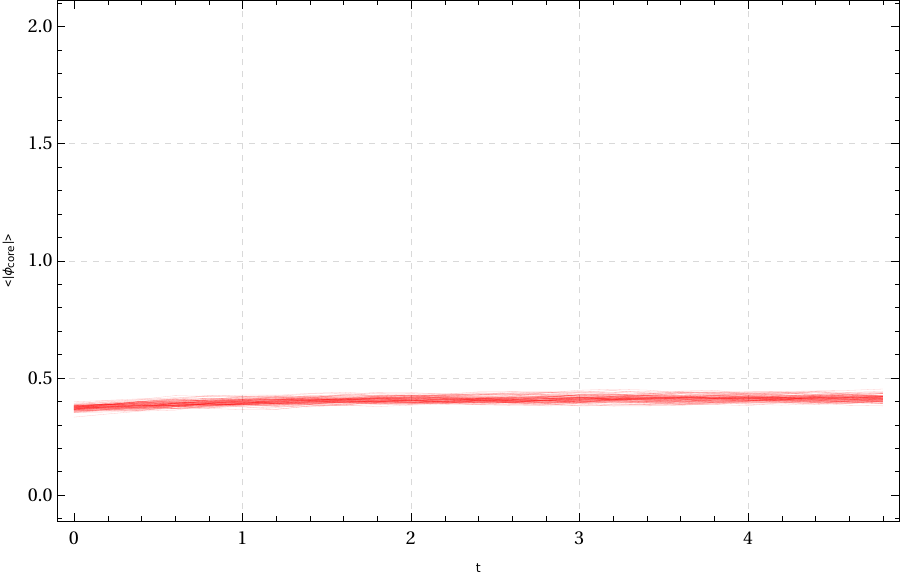}
    \end{minipage}
    \begin{minipage}{0.32\textwidth}
        \includegraphics[width=\textwidth]{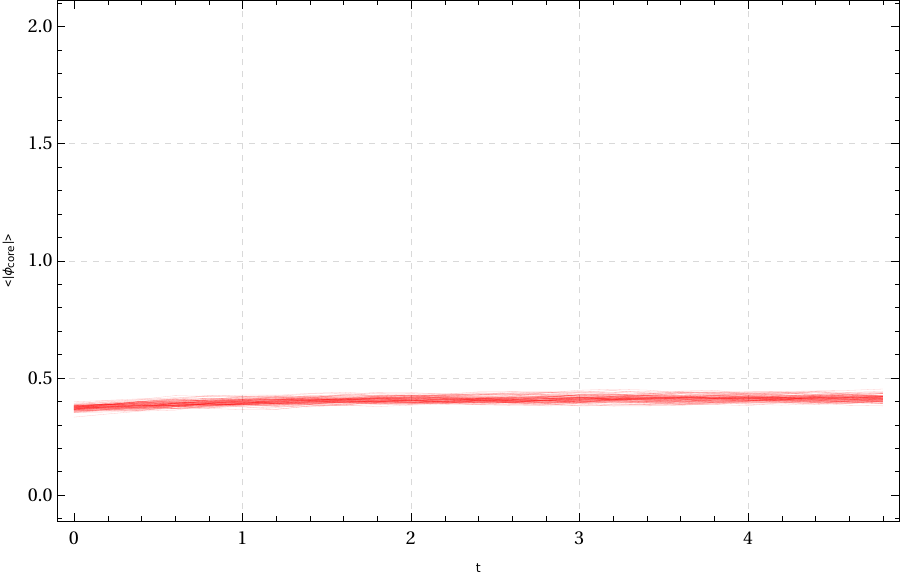}
    \end{minipage}
    \begin{minipage}{0.32\textwidth}
        \includegraphics[width=\textwidth]{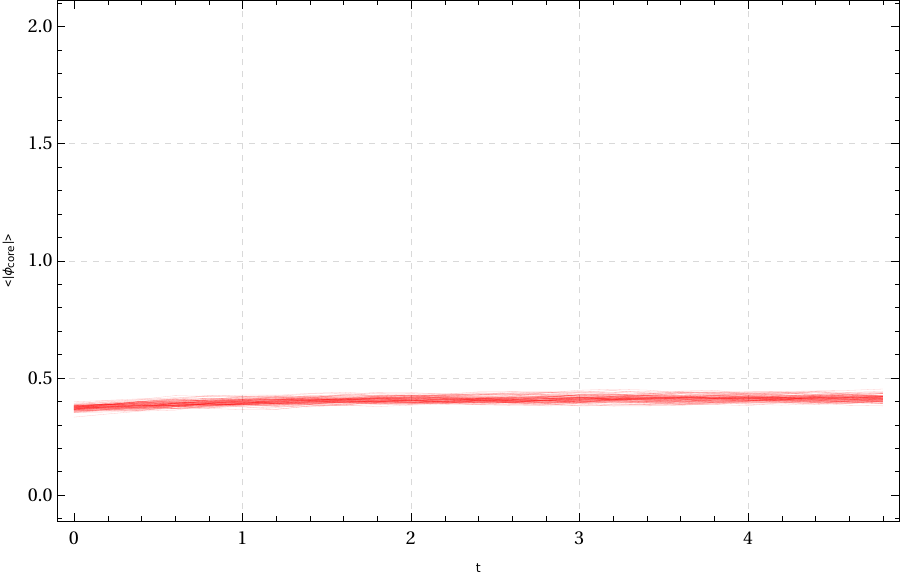}
    \end{minipage}

    \caption{Example evolution of the bubble core, $\langle|\phi_{\rm core}|\rangle$, in the secondary simulations. Blue lines indicate simulations where the core grows toward $\phi_{\rm TV}$, while red lines correspond to cores remaining near the metastable vacuum. The top row shows simulations with initial conditions $(\phi,\pi)$ taken from the primary simulation around a bubble evolving from sub- to super-critical, whereas the bottom row shows control simulations without profile injection. Panels from left to right correspond to increasingly late times for the initial profiles $\phi_0(t)$ and $\pi_0(t)$. As the initial profiles become more critical, more cores evolve to the true vacuum. All control simulations remain near the metastable vacuum; since the two sets share common random numbers (CNR), the observed growth in the top row is entirely due to the injected profiles.}
    \label{fig:Example_Nsub_sims}
\end{figure}

The behaviour of the $N_{\rm sub}$ secondary simulations and the method to extract $p_{\rm B}(\phi_0(t))$ is illustrated in~\cref{fig:Example_Nsub_sims}. Each panel shows $\langle |\phi_{\rm core}|\rangle$ as a function of simulation time, where $\phi_{\rm core}$ is measured within a small radius around the center of the (eventual) super-critical bubble. As time progresses, the core either grows toward the true vacuum (trajectories show in blue) or remains thermally trapped near the false vacuum (shown in red).  

The top row corresponds to simulations with the injected profiles, $\phi_0(t)$ and $\pi_0(t)$, while the bottom row shows control simulations without injection. From left to right, the panels correspond to increasingly late time slices for $\phi_0(t)$, the initial condition for simulation A. The early profile is strongly sub-critical, with only a few cores growing; the intermediate profile is near-critical, with roughly half of the simulations expanding; and the late profile is strongly super-critical, with all cores evolving toward $\phi_{\rm TV}$. This demonstrates that with thermal noise included in the system, the prediction for the evolution of a profile cannot be deterministically determined: even for sub[super]-critical bubbles, the profile can still expand[collapse]. The control simulations in the bottom row remain thermally static, confirming that the growth observed in the top row arises solely from the injected profiles.

Having identified the critical bubble configurations in each of the $N_{\rm primary}$ simulations, we extract their radial profiles by averaging over spherical shells around the bubble center,
\begin{equation}
    \phi(r) \equiv \frac{1}{N(r)} \sum_{i \in [r-\delta r,\,r+\delta r]} \phi_i,
\end{equation}
where $N(r)$ is the number of lattice sites within the shell. This procedure yields smooth, spherically averaged profiles that can be directly compared to theoretical predictions for the critical bubble, providing a direct comparison between the results of our simulations with theory. 
Moreover, by examining the evolution of the committor probability $p_{\rm B}(\phi_0(t))$ alongside the radial profiles, we can quantify how thermal fluctuations influence the growth or collapse of the bubble over time. These combined analyses allow us to characterise both the typical shape of the critical bubble and the statistical nature of nucleation in the presence of stochastic dynamics.

\section{Simulation Results}
\label{sec:4}

\begin{figure}[t]
    \centering
    \includegraphics[width=0.45\linewidth]{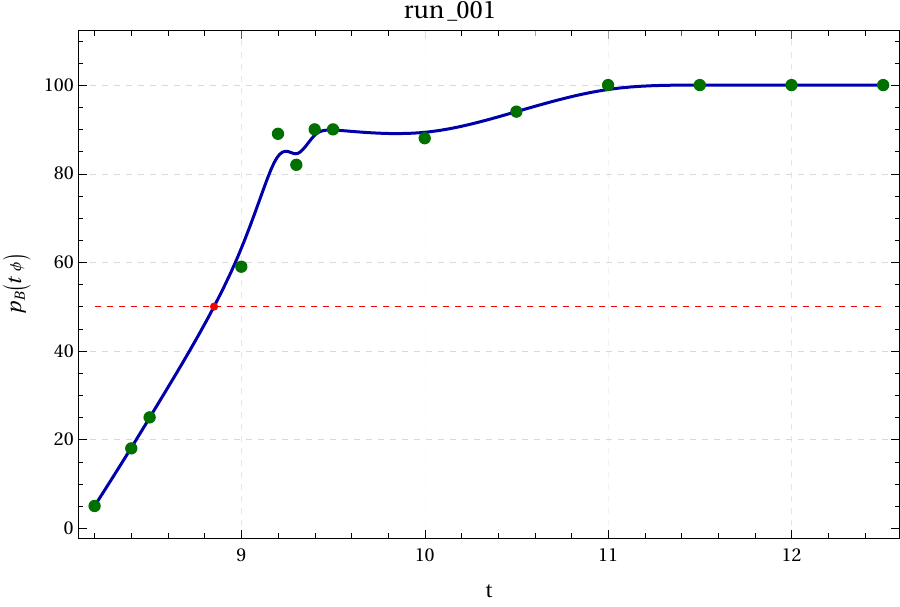}
    \hfill
    \includegraphics[width=0.45\linewidth]{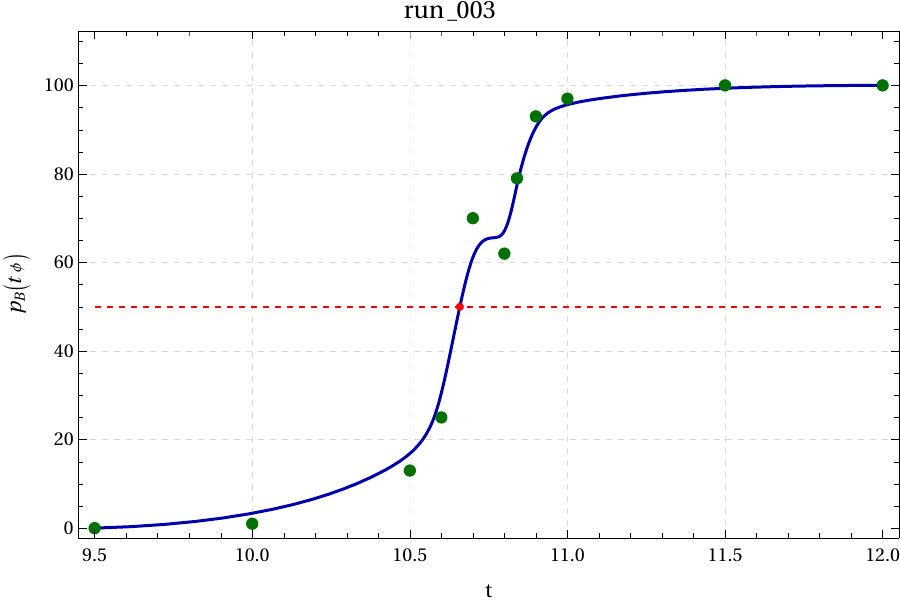}
    \hfill
    \includegraphics[width=0.45\linewidth]{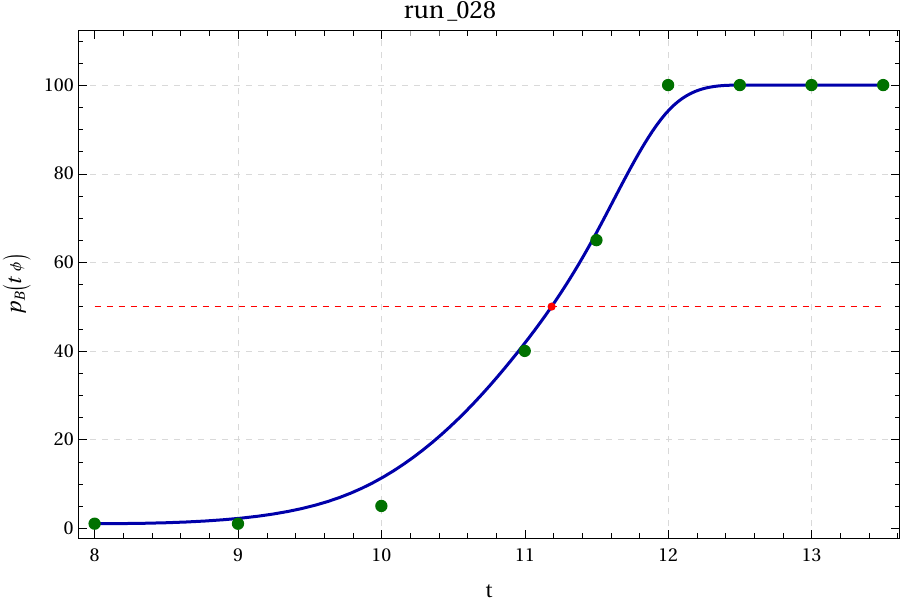}
    \hfill
    \includegraphics[width=0.45\linewidth]{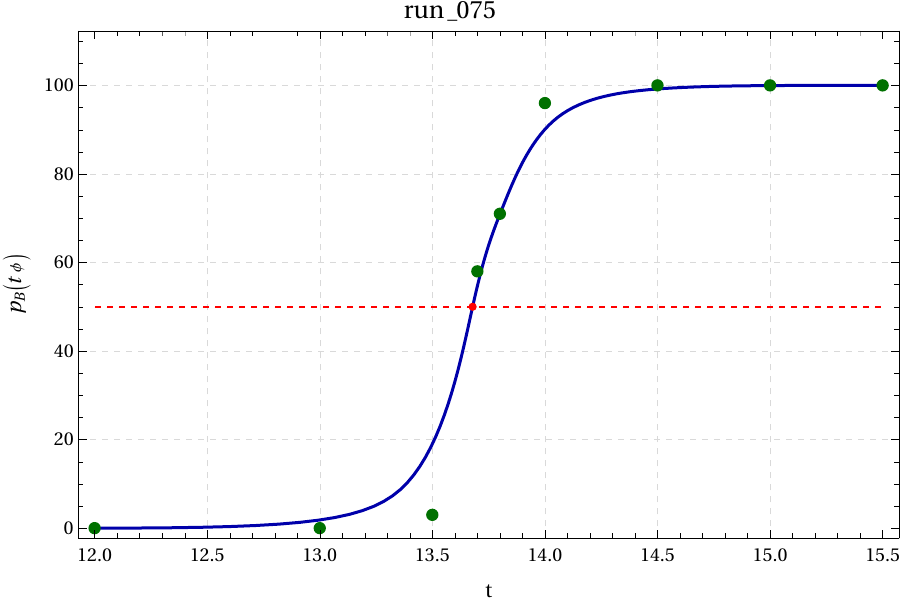}
    \caption{Example plots of the committor probability $p_{\rm B}(\phi_0(t))$ for different primary simulations. In most cases, $p_{\rm B}$ evolves monotonically with time, reflecting the smooth growth of the injected profile toward criticality. The transition from $p_{\rm B} \approx 0$ to $p_{\rm B} \approx 1$ identifies the formation of the critical bubble in each simulation.}
    \label{fig:frac_vs_time_accepted}
\end{figure}

A total of $N_{\rm primary} = 75$ primary simulations were performed. For each primary simulation, the committor probability $p_{\rm B}(t)$ was estimated using roughly $\mathcal{O}(10 \times N_{\rm sub})$ secondary simulations at multiple time slice; that is, $N_{\rm sub}$ simulations each at about 10 different recorded time slices. In each secondary simulation, the corresponding profiles $\phi_0(t)$ and $\pi_0(t)$ from the primary run were injected as an initial condition for the A-type simulations. The resulting data points for $p_{\rm B}(t)$ were then interpolated using a B-spline function to obtain a smooth representation of the committor probability as a function of time. \Cref{fig:frac_vs_time_accepted} shows this extracted behaviour for four representative cases.  

In most instances, $p_{\rm B}(t)$ exhibits largely monotonic growth with time, indicating that the evolution of $\phi_0(t)$ is smooth and that our simulation methodology produces well-defined, consistent outcomes. Minor non-monotonic features are occasionally observed, reflecting the influence of thermal fluctuations: stochastic kicks can temporarily suppress either the field $\phi$ or its momentum $\pi$, causing short-lived decreases in the committor probability.  

Furthermore, the timescale over which the system transitions from sub-critical ($p_{\rm B} \lesssim 0.5$) to super-critical ($p_{\rm B} \gtrsim 0.5$) behaviour varies between simulations. In some cases, the transition occurs very rapidly, over a simulation time span of roughly $t \sim 1$, whereas in others it unfolds more gradually. These variations can be attributed to the stochastic nature of the thermal noise acting on the bubble core: occasionally, a large fluctuation provides a strong push, facilitating rapid critical bubble formation, while in other cases, smaller fluctuations accumulate over time, yielding a slower approach to the critical state.  

Taken together, these observations illustrate that the growth of the bubble core and the associated committor probability are governed by both the deterministic evolution of the profile and the stochastic thermal effects, highlighting the importance of a statistical, rather than purely deterministic, perspective on nucleation dynamics.

\begin{figure}[t]
    \centering
    \includegraphics[width=0.45\linewidth]{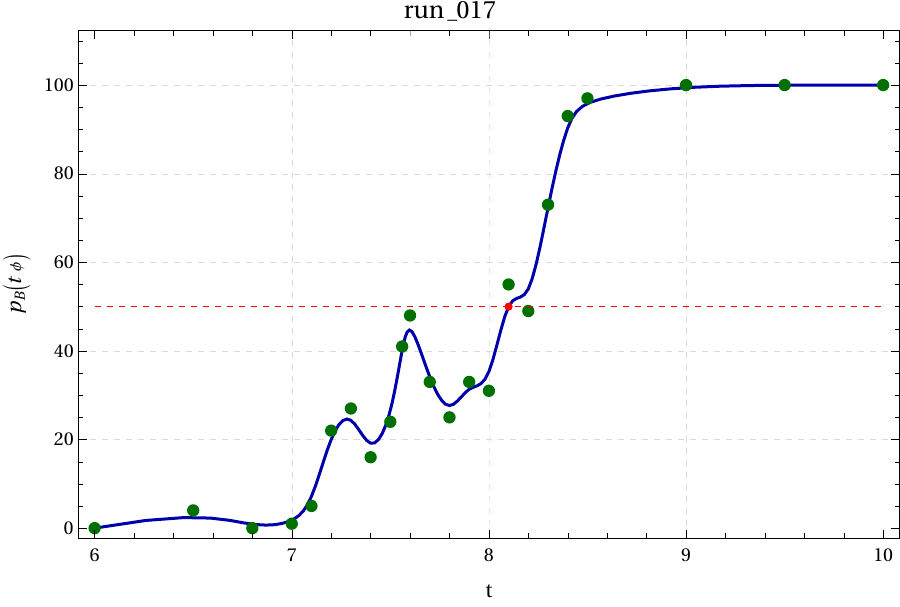}
    \hfill
    \includegraphics[width=0.45\linewidth]{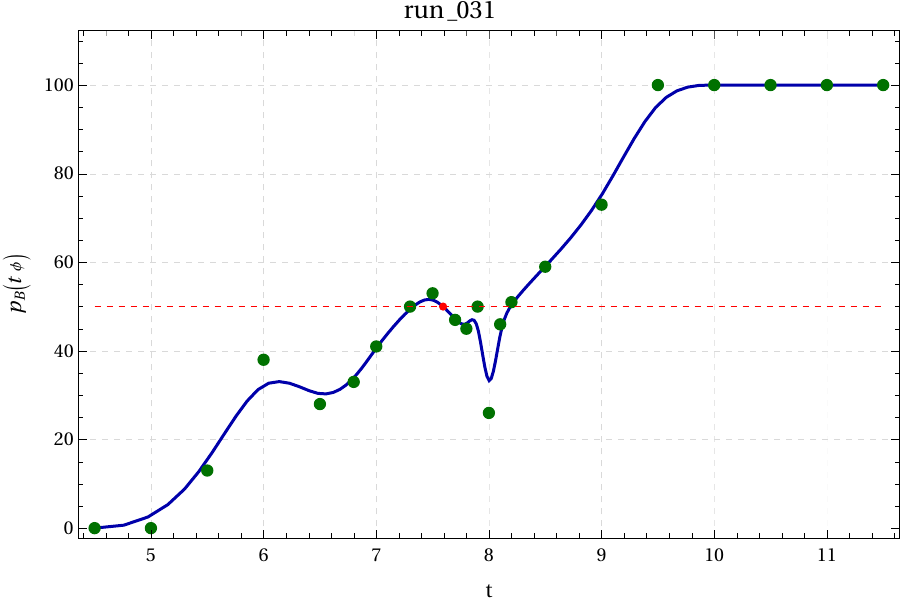}
    \caption{Example of a few runs excluded from the critical bubble analysis. In these cases, multiple time points satisfy the separatrix criterion $p_{\rm B}=1/2$, likely due to thermal fluctuations temporarily suppressing the growing field profile and moving it away from criticality.
}
    \label{fig:frac_vs_time_rejected}
\end{figure}

Interestingly, in a small subset of runs, thermal fluctuations lead to non-monotonic behaviour of the committor probability $p_{\rm B}(t)$, as illustrated in \cref{fig:frac_vs_time_rejected}. In these cases, multiple time points satisfy the criticality condition $p_{\rm B}(t)=1/2$, and the trajectory temporarily moves back and forth across the separatrix. This behaviour is fully consistent with the stochastic dynamics: near the separatrix, thermal fluctuations can transiently suppress or enhance the field configuration, causing it to alternate between configurations more likely to grow and those more likely to collapse.

Such trajectories provide additional information about the structure and width of the transition-state region in configuration space and may be useful for future studies of nucleation dynamics, especially when considering different classes of models. However, the present work focuses on extracting a representative critical bubble profile using trajectories that cross the separatrix in a well-defined manner. For this purpose, we restrict attention to runs with a single, clearly identifiable crossing of the separatrix criterion. Out of the 75 primary simulations performed, 70 satisfy this condition and are used in the construction of the averaged critical bubble profile $\phi^{\rm avg}_{\rm crit}(r)$, while the remaining 5 runs exhibiting multiple crossings are not included in this averaging procedure.

\begin{figure}[t]
    \centering
    \includegraphics[width=0.68\linewidth]{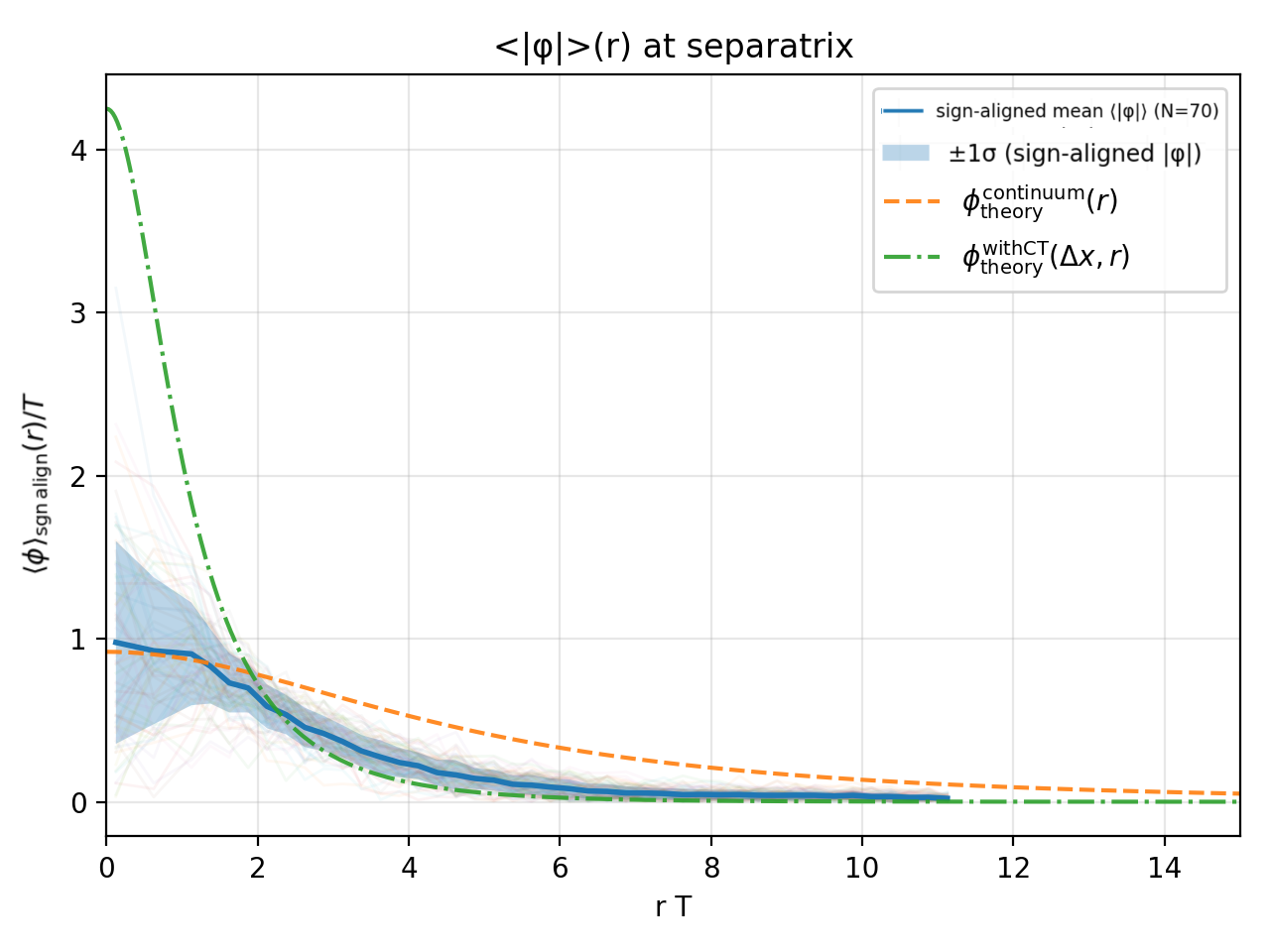}
    \caption{Reconstructed mean critical bubble profile, $\phi_{\rm crit}^{\rm avg}(r)$, from 70 independent lattice runs, with the shaded band indicating the $1\sigma$ spread across simulations. Here $\rm sgn\,align$ denotes a global $Z_2$ sign convention applied after radial averaging: for each run we multiply the radially averaged profile by $s \equiv \rm{sign}(\overline{\phi}_{\rm core})$ so that the core is positive before averaging across runs. This avoids spurious cancellations between runs in the two degenerate vacua and leaves the large-$r$ tail consistent with $\phi(r) \rightarrow 0$. Additionally the $\phi(r)$ behaviour from each run is very faintly drawn overlapping this average result. The dotted green line is the predicted continuum critical bubble including lattice counterterms (used only to demonstrate the large $r$ behaviour), while the red dashed line corresponds to the prediction from the continuum theory. At large $r$, the profile is dominated by the lattice mass counterterm $\delta m^2$ which grows with smaller lattice spacing. The fixed-centre profile agrees well with the predicted core behaviour, whereas recentering produces a spike and does not reproduce the expected $d\phi/dr|_{r=0}=0$ slope. The $1\sigma$ range illustrates the run-to-run variation due to thermal fluctuations and lattice extraction differences.}
    \label{fig:reconstructed_bubble}
\end{figure}

\begin{figure}[t]
    \centering
    \includegraphics[width=0.68\linewidth]{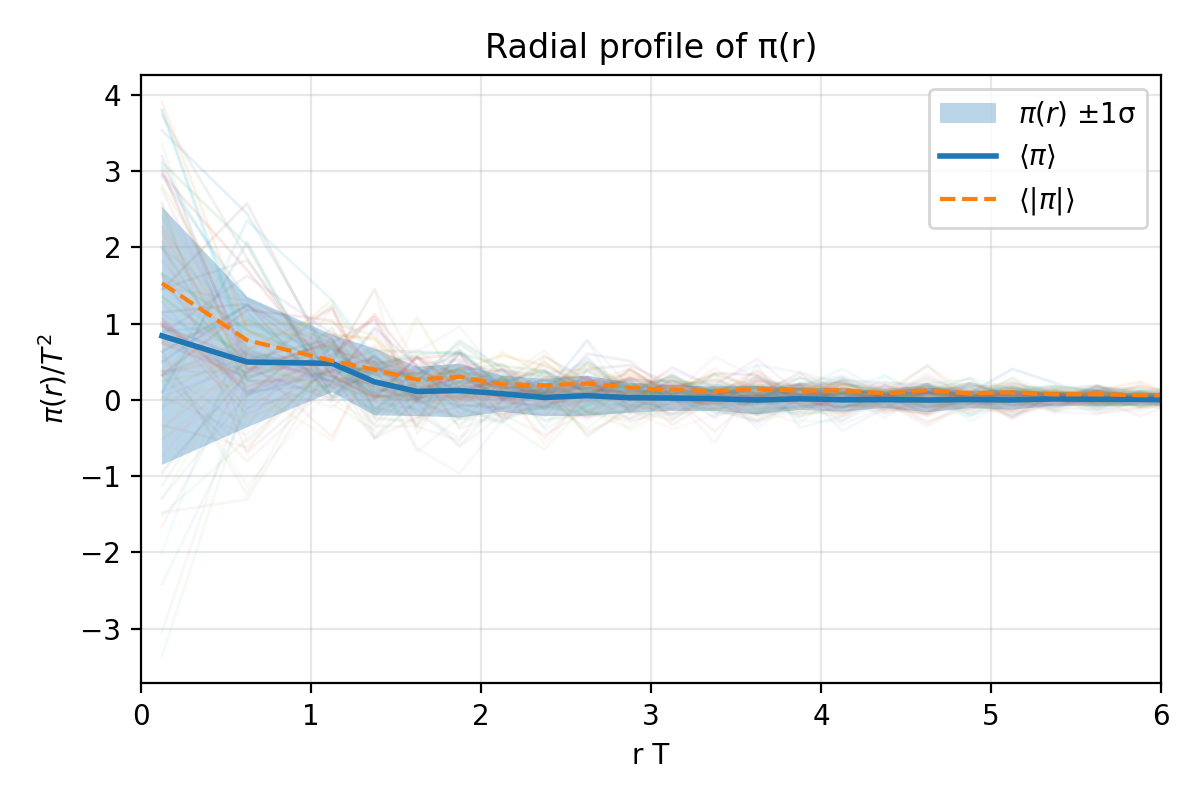}
    \caption{Same figure with~\cref{fig:reconstructed_bubble} but instead plotting $\langle \pi\rangle$ and $\langle |\pi|\rangle$ at the extracted time step of criticality.}
    \label{fig:reconstructed_bubble_mom}
\end{figure}

The reconstructed critical bubble profiles from the 70 primary lattice simulations, including the run-to-run variation, are shown in~\cref{fig:reconstructed_bubble}. This presents the mean profile assuming the centre of the bubble remains fixed in time, with the shaded band indicating the $1\sigma$ variation across independent runs. 
The quantity, $\langle\phi\rangle_{\rm sgn\,align} (r)$, should be understood as taking a radial average of lattice points in each run \emph{without} an absolute value, and then averaging all the profiles obtained from each run making sure each profile is strictly positive in the core region. This is because the two degenerate minima allow for both positive and negative valued critical profiles; if each profile was not first taken to be strictly positive before averaging all obtained runs, spurious cancellations would appear in the ensemble average of $\phi_{\rm crit}(r)$.
Overlapping this are also very faint lines corresponding to the directly reconstructed $\phi(r)$ in each run, which unsurprisingly exhibits some more erratic behaviour compared to the statistical average.
This band provides a measure of the statistical spread due to thermal fluctuations and differences in bubble extraction between runs. Even with these variations, the core of the un-recentered profile aligns well with the predicted continuum critical bubble including lattice counterterms. 

Far from the bubble centre, the field approaches the false vacuum and the equation of motion linearises. In this regime the profile satisfies
\begin{equation}
\frac{d^2 \phi}{dr^2} + \frac{2}{r}\frac{d\phi}{dr}
\simeq
m_{\rm eff}^2\,\phi,
\end{equation}
with the universal solution
\begin{equation}
\phi(r)\propto \frac{e^{-m_{\rm eff} r}}{r}.
\end{equation}
Here the effective mass includes the lattice counterterm,
\begin{equation}
m_{\rm eff}^2 = m^2 + \delta m^2,
\end{equation}
and therefore the decay length of the tail is directly sensitive to lattice renormalisation effects. Since the lattice mass counterterm provides a parametrically large contribution to $m_{\rm eff}^2$ in this case\footnote{As \cref{Eq:S3overT_simulation} shows, this is almost required as a sufficiently small input nucleation rate for reasonable simulation time requires choosing a large $\lambda$ and sufficiently small $m^2$: so $\delta m^2 \sim \lambda/a > m^2$ unless lattice spacings are unacceptably large}, it plays a significant role in determining the asymptotic falloff of the profile.

By contrast, near the bubble core where $\phi \sim \phi_{\rm vev}$, the dynamics are governed by the full nonlinear structure of the potential. In this region the quartic and sextic terms are comparable,
\begin{equation}
\lambda \phi^3 \sim \epsilon \phi^5,
\end{equation}
and the profile shape is controlled by the balance between nonlinear interactions rather than by the quadratic mass term alone. 

Finally, the contribution of the tail region to the action density,
\begin{equation}
\mathcal{I}(r) = 4\pi r^2 \left[ \frac{1}{2} (\partial_r \phi)^2 + V(\phi)\right],
\end{equation}
is exponentially suppressed due to the small field amplitude. As a result, while lattice renormalisation effects significantly influence the asymptotic decay of the profile, they have only a minor impact on the total action and the extracted nucleation rate, which are dominated by the bubble wall region.

A recentering procedure within a three-lattice-site neighbourhood was tested, again with a $1\sigma$ band. The recentering procedure involves shifting the assumed bubble centre to the lattice site within a small local neighbourhood that has the largest field value, with the goal of compensating for possible small drifts of the bubble core due to lattice discretisation or thermal fluctuations during evolution. While this can help align profiles across independent runs and reduce apparent misalignment, it also produced a pronounced spike in the core. This peak does not reproduce the expected $d\phi/dr|_{r=0} = 0$ behaviour, suggesting that recentering may be simply selecting a site near the true core which received a large enough fluctuation to be the largest lattice site, rather than the actual centre of the bubble. From this it we conclude that bubble drift does not occur within these simulations.

Overall, the $1\sigma$ bands illustrate that the reconstructed critical bubble is statistically well-defined. It is quite remarkable that, in this simulation, the core of the extracted profile seems to match well what is predicted by theory. 
Further simulations are required with other parameters and potentials to see if this behaviour can be consistently reproduced or not. This is left to future work simply due to the large computational time taken in producing these results.

An advantage of using the statistical committor-based definition of the critical bubble is that it allows not only the field profile $\phi(r)$ but also the associated radial momentum properties to be investigated. \Cref{fig:reconstructed_bubble_mom} illustrates this explicitly. At the point of criticality, the bubble core develops a nonzero momentum density, $\pi(r)$, indicating that the configuration is not a perfectly static saddle-point in field space but instead carries dynamical structure. This behaviour is physically reasonable: for a bubble to successfully compete with plasma noise and fluctuations in the environment, it may require the buildup of a localised momentum “kick” in its interior that enables it to cross the separatrix between collapse and growth. In other words, the momentum core could be viewed as part of the minimal dynamical ingredients needed for the bubble to seed expansion.
The behaviour seen in \cref{fig:reconstructed_bubble_mom} closely parallels the phase-space trajectories shown in fig.~2 of Ref.~\cite{Hirvonen:2024rfg}. 
In that work, trajectories of the bubble were projected onto the $(\phi,\pi)$ plane: configurations that recollapsed traced the ``deflecting'' branch while those that expanded followed the ``nucleating''.
The overdamped critical bubble lies exactly on the $\pi=0$ separatrix between these two families. 
Our ensemble of statistical critical bubbles at finite~$\eta$ occupies the neighbourhood of this separatrix—effectively the region between the deflecting and nucleating trajectories—consistent with the small but nonzero momentum densities observed in the simulations.

It is worth stressing that in the standard semiclassical treatment of nucleation the critical bubble is obtained as a static solution of the Euclidean field equations, and hence is assumed to satisfy $\pi(r)=0$ everywhere (see e.g.~\cite{Coleman:1977py,Langer:1969bc}). The appearance of nonzero momentum in our lattice results therefore reflects the inherently stochastic and thermal nature of the real-time simulations, where noise and fluctuations can excite dynamical modes even at the critical configuration.

Using the extracted critical bubble profile, we can estimate the value of $S_3$.  
We adopt a clean discrete definition of the three–dimensional Euclidean action,
\begin{equation}
S_3 \;\approx\; 4\pi \sum_{i = r_{\rm start}}^{r_{\rm max}} \Delta r \; r_i^2 
\left[
\tfrac{1}{2} \left(\frac{\phi_{i+1} - \phi_{i-1}}{2 \Delta r}\right)^2
+ V(\phi_i) - V(\phi_{\rm false})
\right] \,,
\label{eq:S3_discrete}
\end{equation}
where several practical implementation details are required. The radial derivative 
$\partial_{r}\phi$ is evaluated using central finite differences, with forward or backward 
differences applied at the ends of the lattice. The radial integral itself is performed with 
the trapezoidal rule on integer lattice bins $r=0,1,2,\dots$. To mitigate the large 
fluctuations that occur very close to the nominal bubble centre, where the number of 
lattice points per spherical shell is small and the thermal noise source $\xi$ has a large  relative impact, the integration is started from a nonzero radius, $r_{\rm start}$ to avoid these large fluctuations.
For the potential we use the continuum form, deliberately excluding the lattice 
counterterms. 

The extracted values for the three-dimensional Euclidean action $S_3$, obtained from the reconstructed bubble profiles in~\cref{fig:reconstructed_bubble}, are  
\begin{equation}
    S_3 = 6.4 \pm 1.72 \,,
\end{equation}
using the left-hand profile, and  
\begin{equation}
    S_3 = 9.19 \pm 1.85 \,,
\end{equation}
using the right-hand profile.  
These results can be compared to the theoretical predictions listed in~\cref{tab:c_H}. Overall, the agreement is quite good. The smaller tail of $\phi_{\rm crit}^{\rm avg}(r)$ obtained compared to the prediction has only a small impact on the $S_3$ extracted from the lattice profiles. This is expected as the contribution to the total free energy from the bubble periphery is subdominant relative to the core, which dominates the integral in~\cref{eq:S3_discrete}. This observation confirms that the lattice-extracted profiles capture the essential physics of the critical bubble, even if deviations in the tail exist due to the presence of finite lattice counterterms and thermal fluctuations.

\section{Conclusion}
\label{sec:5}

In this work, we have developed a fully statistical framework for identifying critical bubbles in thermally fluctuating quantum field theories. Building on the concept of the committor probability from chemical reaction theory, we defined the critical bubble as the field configuration for which the probability to expand to the true vacuum equals the probability to collapse back to the false vacuum, $p_B = 1/2$. This probabilistic definition naturally generalises the conventional bounce solution and separatrix picture, reducing to it in the noiseless limit while accommodating the stochasticity introduced by finite-temperature fluctuations.

Using a combination of large-scale primary lattice simulations and targeted secondary subgrid simulations, we demonstrated a practical procedure to extract this statistical critical bubble. The primary simulations serve to identify candidate bubble configurations, while the secondary simulations, performed with common random numbers, allow for an unambiguous measurement of the committor probability. By interpolating between time slices, we can pinpoint the profile that is statistically critical, distinguishing subcritical and supercritical configurations. 
We find very good consistent behaviour for the committor probability, $p_{\rm B}$, as it evolves in time for most simulations highlighting the consistency of the method. We can also use this to examine the behaviour of $\pi(r)$ at the extracted moment of criticality.

Our results highlight several important points. First, thermal noise can modify the deterministic picture: (i) configurations that would be supercritical in a noiseless evolution may occasionally induce a reduction in $p_{\rm B}(t)$, likely caused by local thermal fluctuations, (ii) similarly subcritical configurations may sometimes nucleate due to stochastic fluctuations ($p_{\rm B}(t)$ sometimes grows faster or slower), (iii) it is possible in rare cases for the behaviour of $p_{\rm B}$ to oscillate around the criticality boundary and, (iv) the statistical approach provides a robust method to check analytic estimates of the critical bubble, offering a simulation-based validation of semi-analytic bounce calculations. Finally, the framework is flexible: it can accommodate inhomogeneous, asymmetric, or otherwise non-ideal initial conditions, which are challenging to capture in traditional analytic treatments.

Looking forward, this approach opens several avenues for future work. Currently we have implemented a very na\"ive definition of bubble commitment in a given simulation: by simulating for a time $t_f$ and observing the behaviour of the (eventual) core. More robust or computationally faster methods of determining whether a given simulation has committed to the true vacuum or not, may not only improve computational efficiency, but improve the derived $p_{\rm B}$. We find good agreement between the our extracted critical profile and that from theory, but this needs to be more rigorously studied with other parameters and potentials to determine its consistency, which will take some computational time. As this is a purely statistical criterion, it may be interesting to study the nucleation rates in more complex potentials or multi-field systems where analytic bounces are difficult to obtain. Finally, the statistical critical bubble concept could serve as a benchmark for improving approximate semi-analytic nucleation rate formulas or for guiding effective field theory constructions in finite-temperature environments.

In summary, by re-framing the critical bubble as a statistical dividing surface in configuration space, we provide a concrete, numerically accessible idea to study first-order phase transitions in realistic thermal settings, bridging the gap between idealised bounce solutions and the stochastic dynamics of fluctuating fields.

\section*{Acknowledgement}

TPD thanks Chang Sub Shin and Tae Hyun Jung for valuable feedback on the manuscript. TPD is supported by KIAS Individual Grants under Grant No. PG084101 at the Korea Institute for Advanced Study and thanks Suro Kim for continued discussions throughout the course of this project.

\appendix
\section{Lattice Discretisation of the Langevin Equation}
\label{app:A}

In order to solve~\cref{eq:sec2langevin}, we discretise the dimensionally reduced 3-dimensional theory, replacing \(dx \rightarrow a\) and \(dt \rightarrow \Delta t\), and numerically evolve the field at each lattice site. For the spatial derivatives, we adopt an improved discretisation of the Laplacian with $\mathcal{O}(a^4)$ accuracy:
\begin{equation}
\nabla^2 \phi(\mathbf{x}) = \frac{1}{12 a^2} \sum_i \Big[ -\phi(\mathbf{x}-2\hat{i}) + 16 \phi(\mathbf{x}-\hat{i}) - 30 \phi(\mathbf{x}) + 16 \phi(\mathbf{x}+\hat{i}) - \phi(\mathbf{x}+2\hat{i}) \Big],
\end{equation}
where \(\hat{i}\) denotes the unit vector along each lattice direction.

The non-smooth noise term in~\cref{eq:sec2langevin} is discretised by replacing the continuum correlator
\begin{equation}
\langle \xi(\mathbf{x},t) \xi(\mathbf{x}',t') \rangle_T = D \, \delta(t-t') \delta^3(\mathbf{x}-\mathbf{x}')
\end{equation}
with its lattice analogue
\begin{equation}
\langle \xi(\mathbf{x}_i, t_a) \xi(\mathbf{x}_j, t_b) \rangle = \frac{D}{(\Delta t) a^3} \delta_{ij} \delta_{ab}.
\end{equation}
This can be simulated at each lattice point using independent Gaussian-normal random variables
\begin{equation}
\xi(\mathbf{x}_i, t_a) = \sqrt{\frac{D}{\Delta t \, a^3}} \, \mathcal{G}_{i,a}.
\end{equation}
These choices ensure the correct variance of \(\xi\) when mapping from the continuum to the lattice, although the strong convergence of the numerical scheme is limited to \(\mathcal{O}(a)\)~\cite{telatovich2020strongconvergenceoperatorsplittingmethods}.\footnote{The independent Gaussian noise ignores correlations at scales below the lattice spacing. For very fine features, e.g., near the bubble core, this may lead to slightly nonphysical sharp gradients.}

\medskip

When discretising a continuum field theory with thermal fluctuations, ultraviolet (UV) divergences arise due to the finite lattice spacing \((\Lambda_{\mathrm{UV}} \propto a^{-1})\). To properly match the equilibrium continuum theory, lattice renormalisation counterterms must be included. These can be derived using lattice perturbation theory and dimensional analysis. Denoting a generic coupling as \(X\), the lattice counterpart is~\cite{Gould:2024chm}  
\begin{equation}
X_{\rm{lat}} = X_{\rm{cont}} + \delta X,
\end{equation}
for e.g. \(X = m, \lambda, \epsilon\) which correspond to mass, quartic, and sextic couplings, respectively.  

The lattice-discretised simulation of~\cref{eq:sec2langevin} then involves the replacements
\begin{align}
V_T(\phi) &\rightarrow V_3 (\phi) = \frac{1}{2} Z_\phi Z_m (m_3^2 + \delta m_3^2) \phi^2 - \frac{1}{4!} Z_\phi^2 (\lambda_3 - \delta \lambda_3) \phi^4 + \frac{1}{6!} Z_\phi^3 (\epsilon_3 + \delta \epsilon_3) \phi^6, \\
\nabla^2 \phi &\rightarrow Z_\phi \nabla^2 \phi.
\end{align}

For $\phi^4$ theories, the lattice renormalisation terms have been computed up to $\mathcal{O}(a^2)$~\cite{Moore:2001vf, Arnold:2001ir, Sun:2002cc}. In our case, we include additional contributions from $\epsilon_3$, although these are small when $\epsilon_3 \ll 1$. The leading-order contributions are
% \begin{align}
% \delta m_3^2 &= -\frac{\Sigma \lambda_3}{8\pi a} + \mathcal{O}(\lambda_3^2 a^0),\\
% \delta \lambda_3 &= -\frac{\Sigma \epsilon_3}{8\pi a} + \mathcal{O}(\epsilon_3^2 a^0) + \mathcal{O}(\lambda_3^2 a),\\
% \delta \epsilon_3 &= 0 + \mathcal{O}(\epsilon_3^2 a),\\
% Z_\phi &= 1 + \mathcal{O}(a^2),\\
% Z_m &= 1 + \mathcal{O}(a),
% \end{align}
\begin{align}
\delta m_3^2
&=
-\frac{\Sigma \lambda_3}{8\pi a}
+
\frac{\lambda_3^2}{16\pi^2}
\left(
\log\!\left(\frac{6}{a |\lambda_3|}\right)
+ C_3 - \Sigma x_i
\right)
+ \mathcal{O}(\lambda_3^3 a),
\\
\delta \lambda_3
&=
-\frac{\Sigma \epsilon_3}{8\pi a}
+
\frac{3 x_i \lambda_3^2 a}{8\pi}
+
\frac{\lambda_3^2 a^2}{64\pi^3}
\left(
\frac{3}{4}x_i^2 - 3C_1 - \frac{C_2}{3}
\right)
+ \mathcal{O}(\lambda_3^3 a^3),
\\
\delta \epsilon_3
&=
\frac{5 \epsilon_3^2}{48\pi^2}
\left(
\log\!\left(\frac{6}{a |\lambda_3|}\right)
+ C_3 - \Sigma x_i
\right)
+ \mathcal{O}(\epsilon_3^3 a),
\\
Z_\phi
&=
1
+
\frac{C_2 \lambda_3^2 a^2}{96\pi^2}
+
\mathcal{O}(\lambda_3^3 a^3)
\end{align}
where $\Sigma$ is a renormalisation constant that depends on the choice of $\nabla^2$ operator. The values used in our simulations are summarised in~\cite{Arnold:2001ir, Sun:2002cc, Gould:2024chm}.

We note that, for the benchmark considered in~\cref{tab:c_H}, where
\(
\epsilon = T^2/100\text{ and }\lambda=2, 
\)
the majority of counterterm contributions proportional to the sextic coupling produce numerically small shifts.

An additional contribution arises at two-loop order from the ``double-tadpole'' (flower) diagram involving the sextic interaction, in which four of the six legs are contracted to form two tadpole loops. This topology has been discussed, for example, in~\cite{Croon:2020cgk} in the analogous SMEFT context. In the present normalization, this diagram generates an additional mass correction
\begin{equation}
\delta m_3^2\Big|_{\epsilon_3}
=
-\frac{\epsilon_3}{2}
\left(\frac{\Sigma}{8\pi a}\right)^2.
\end{equation}

Although this contribution diverges more strongly with lattice spacing than the leading one-loop mass counterterm proportional to $\lambda_3$, its numerical impact depends on the physical lattice spacing used in the simulations.

In the simulations performed here, the lattice spacing and time step are fixed in physical units to
\begin{equation}
aT = 0.75 \quad\text{and} \quad \Delta t / T = 0.005
\end{equation}

Using the benchmark parameters, we estimate the relative size of this correction compared to the leading mass counterterm to be
\begin{equation}
\frac{\delta m_3^2|_{\epsilon_3}}
     {\delta m_3^2|_{\lambda_3}}
\simeq 4\times 10^{-4},
\end{equation}
indicating that the sextic-induced contribution produces a negligible shift in the effective potential for the lattice parameters used here.

More generally, although this contribution is formally more ultraviolet-divergent, it remains numerically subdominant at fixed physical lattice spacing in the present regime. Its inclusion would correspond primarily to a small renormalisation of the effective mass parameter and would not significantly affect the conclusions of this work however we stress that if such simulations were to be performed with relatively smaller[larger] values of $\lambda_3[\epsilon_3]$, this terms contributions may quickly begin to dominate and should not be ignored.

\bibliographystyle{JHEP}
{\footnotesize
\bibliography{biblio}}

\end{document}